\documentclass[12pt]{article}
\usepackage{amssymb,amsmath,graphicx,epsfig}
\newcommand{\eldpc}[3]{\rm{LDPC}(#1, #2, #3)}

\newcommand{\ledge}{\lambda}           % left degree distribution-edge perspective
\newcommand{\redge}{\rho}              % right degree distribution-edge perspective
\def\prob{{\mbox{\rm P}}}

\begin{document}

\renewcommand{\textfraction}{0}
\def\eprob{\epsilon}
\def\bec{{\rm BEC}(\epsilon)}
\def\block{{\rm P}_{\rm B}}
\def\bit{{\rm P}_{\rm b}}
\def\eit{\epsilon^*}
\newtheorem{lemma}{Lemma}
\newtheorem{conjecture}{Conjecture}
\def\ux{\underline{x}}
\def\uz{\underline{z}}

\title{Finite-Length Scaling and Finite-Length Shift for
Low-Density Parity-Check Codes} 
\author{
\normalsize Abdelaziz Amraoui \\
\small EPFL (Lausanne), CH-1015\\
\small {\tt abdelaziz.amraoui@epfl.ch}\\
\and
\normalsize Andrea Montanari\\ 
\small LPTENS (UMR 8549, CNRS et  ENS) \\[-5pt] 
\small 24, rue Lhomond, 75231  \\[-5pt] 
\small  Paris CEDEX 05, France \\
[-5pt] \small {\tt montanar@lpt.ens.fr}
\and
\normalsize Tom Richardson\\
\small Flarion Technologies\\
\small Bedminster, NJ,  USA-07921\\
\small {\tt richardson@flarion.com}\\
\and
\normalsize R{\"{u}}diger Urbanke\\
\small EPFL (Lausanne), CH-1015\\
\small {\tt rudiger.urbanke@epfl.ch}\\
 }
\date{}
\maketitle
\thispagestyle{empty}
\begin{abstract}
Consider communication over the binary erasure channel
$\bec$ using random low-density parity-check codes with 
finite-blocklength $n$ from  `standard' ensembles.
We show that  large error events is conveniently
described within a {\em scaling theory}, and explain how to estimate
heuristically their effect.
Among  other quantities, we consider  the finite length threshold
$\eit(n)$, defined by requiring a block error probability $\block = 1/2$. 
For ensembles
with minimum variable degree larger than two, the following expression is 
argued to hold
\begin{eqnarray*}
\eit(n) = \eit -\eit_1\, n^{-2/3} +\Theta(n^{-1})\, ,
\end{eqnarray*}
with a calculable {\em shift} parameter $\eit_1>0$.
\end{abstract} 
\normalsize

\section{Introduction}

Assessing the performances of finite-blocklength iterative coding systems
is an important open issue in modern coding theory. A particular case of 
such a task consists in the study of low-density parity-check code
(LDPC) ensembles, when used for communicating over the binary erasure channel
$\bec$. A consistent effort has been devoted to this case,  with
the hope of a  positive feedback on the general 
problem~\cite{DPRTU01,RSU02bis,RSU02,RiU02,ZaO02}. 
Some lessons can be drawn from the results obtained so far:
\vspace{0.2cm}

{\bf Approximate!} While density evolution (DE) provides exact thresholds
in the large blocklength limit, there is little hope to 
 compute exact performances (bit or block error rates 
$\bit$ and $\block$) at finite blocklength $n$. 
For the $\bec$, $\bit$ and $\block$ are determined by 
a set of recursions \cite{DPRTU01} 
whose evaluation has  complexity $\Theta(n^{\kappa})$. 
However the exponent $\kappa$ grows with the irregularity
of the ensemble (more precisely, with the number of probabilistically
inequivalent types of node in the Tanner graph). Given the large degree 
of irregularity necessary for approaching capacity, an exact calculation 
becomes prohibitive already for moderate blocklengths.
The situation can unlikely be simpler for more general channel models.

It is therefore crucial to develop approximate estimates of 
finite-length performances.
\vspace{0.2cm}

{\bf Small error events.} Consider, for the sake of simplicity, communication 
over $\bec$ using an LDPC ensemble. Below the threshold 
$\eit$ for iterative decoding, the typical size of error 
events (the number of erased bits after decoding) is of order $1$. 
Above threshold, the same size is of order $n$ (the bit error probability is 
finite). One can regard the failure of iterative decoding at $\eit$
as due to the divergence (on the $\Theta(1)$ scale) of the size of 
typical error events.

The probability of small error events is readily evaluated through the 
union bound. For a code with minimum variable degree $l_{\rm min}$,
the expected number of  error events involving
$E$ bits is $\Theta(n^{E-\lceil El_{\rm min}/2\rceil}\eprob^{E})$,
as long as $E$ is kept finite in the $n\to\infty$ limit.
If $l_{\rm min}>2$, this quantity decreases by a factor $n$ (or more)
each time $E$ increases by one. This suggests that only very small
error events have a non-negligible probability: 
their contribution can be computed
exactly~\cite{RSU02bis} and compares favorably with recursive calculations or
numerical simulations. Moreover, in the vast majority of elements from the
ensemble, such error events are strictly absent.

If $l_{\rm min}=2$, the dominating error events involve uniquely
degree 2 variable nodes, and have the topology of cycles in the Tanner 
graph. The number of such structures is $\Theta(n^{2E})$: one can chose
both the variables and the check nodes involved. Their probability
is $\Theta(\Lambda_2^E\epsilon^En^{-2E})$, where we denoted by $\Lambda_l$
the fraction of variable nodes having degree $l$. In fact: 
the variables must be erased  (which explains the factor $\eprob^{E}$); 
they must have degree 2 (factor $\Lambda_2^E$);
and they must be connected to the previously  chosen check nodes 
(factor $n^{-2E}$). Unlike in the case $l_{\rm min}>2$,  the resulting typical 
size depends upon $\eprob$. Detailed calculations can be carried on for 
cycle ensembles: one finds $E_{\rm typ}\sim |\eit-\eprob|^{-1}$ as
$\eprob\uparrow\eit$. The divergence of error events size `drives' 
the failure of iterative decoding above $\eit$.
\vspace{0.2cm}

{\bf Large error events.} Computing the probability of small
(i.e. finite in the $n\to\infty$ limit) error events is a 
conceptually straightforward task. While in the case $l_{\rm min}>2$
one has do the computation for just a few small structures,
if degree-two nodes are present, an infinite number of contributions
must be summed over. In both cases, this approach yields a simple and 
accurate description of the error probability in the noise regime
for which $\bit =\Theta(n^{1-\lceil l_{\rm min}/2\rceil})$.
This is the  so-called {\em error-floor} region.

What about the {\em waterfall} region? The description in terms of finite-size
error events cannot account for this regime. Take for instance the 
case $l_{\rm min}>2$. As long as the error size $E$ is finite, 
its probability decreases rapidly with $E$. The breakdown of iterative 
decoding at $\eit$ can therefore be ascribed uniquely 
to error events whose size diverges with $n$. 
Analogously, for cycle ensembles the typical size of finite error
events diverges at $\eit$. This conclusion can be extended to general
irregular ensembles: in order to describe the waterfall region,
large error events have to be taken into account.

This remark implies several theoretical problems. First of all, no 
enumeration of all the configurations (by this we mean stopping sets in the
$\bec$ case, and any suitable generalization for other channels)
responsible for errors is possible in the waterfall regime.
In fact, it is likely that the number of relevant `topologies' diverges 
with the blocklength. Second, we cannot think of the set of wrongly decoded 
bits as the union of several small `error events', which are 
probabilistically independent. In more practical terms: the union bound is
not a reliable tool in this regime.

Yet, as optimized ensembles approach capacity, controlling the waterfall 
region is of utmost interest. We developed an approach to this problem
for the $\bec$ case, which yields extremely satisfactory results. The 
methods are complementary to the stopping-sets analysis and build 
upon the description of iterative decoding by Luby et 
al.~\cite{LMSSS97,LMSS01}.
Although the extension to general channel models is likely to require a 
considerable effort, one can learn a general lesson about which 
kind of characterization can be hoped for in the waterfall regime.

Consider, for the sake of simplicity the case of $l_{\rm min}>2$.
We find that there exists a non-negative constant $\nu$ 
and some non-negative function $f(z)$ so that
\begin{eqnarray}
\lim_{ n \rightarrow \infty} 
\prob_{\rm B}(n, \epsilon_n) = f(z)\, ,\label{equ:generalfss}
\end{eqnarray}
where the $n\to\infty$ limit is taken by keeping 
$n^{\frac{1}{\nu}}(\epsilon^*-\epsilon_n) = z$ fixed.
In other words, if one plots $\prob_{\rm B}(n, \epsilon)$
as a function of $z=n^{\frac{1}{\nu}} (\epsilon^*-\epsilon)$
then, for increasing $n$ these finite-length curves
are expected to converge to some function $f(z)$. The function
$f(z)$ decreases smoothly from $1$ to $0$ as its argument changes from 
$-\infty$ to $+\infty$.
This means that  all finite-length curves are, to first order, 
scaled versions of some {\em mother} curve $f(z)$.
It might be helpful to think of the
threshold $\epsilon^{*}$ as the zero order term in
a Taylor series. Then the above scaling, if correct, represents
the first order term.
In fact, one can even refine the analysis to include higher order terms and 
write
\begin{eqnarray}
\prob_{\rm B}(n, \epsilon) & = & f(z) + n^{-\omega} g(z) + o(n^{-\omega}),
\label{equ:generalreffss}
\end{eqnarray}
where $\omega$ is some positive real number and $g(z)$ is the 
second order correction term. 

For ensembles with $l_{\rm min}=2$ (and in particular, for ensembles
whose threshold is fixed by the local stability condition) 
Eq.~(\ref{equ:generalfss}) must be properly generalized. We refer to
Ref.~\cite{AMRU04b} for an example.

\begin{figure}[htp]
\begin{center}
\setlength{\unitlength}{0.8bp}%
\begin{picture}(0,0)
\includegraphics[scale=0.8]{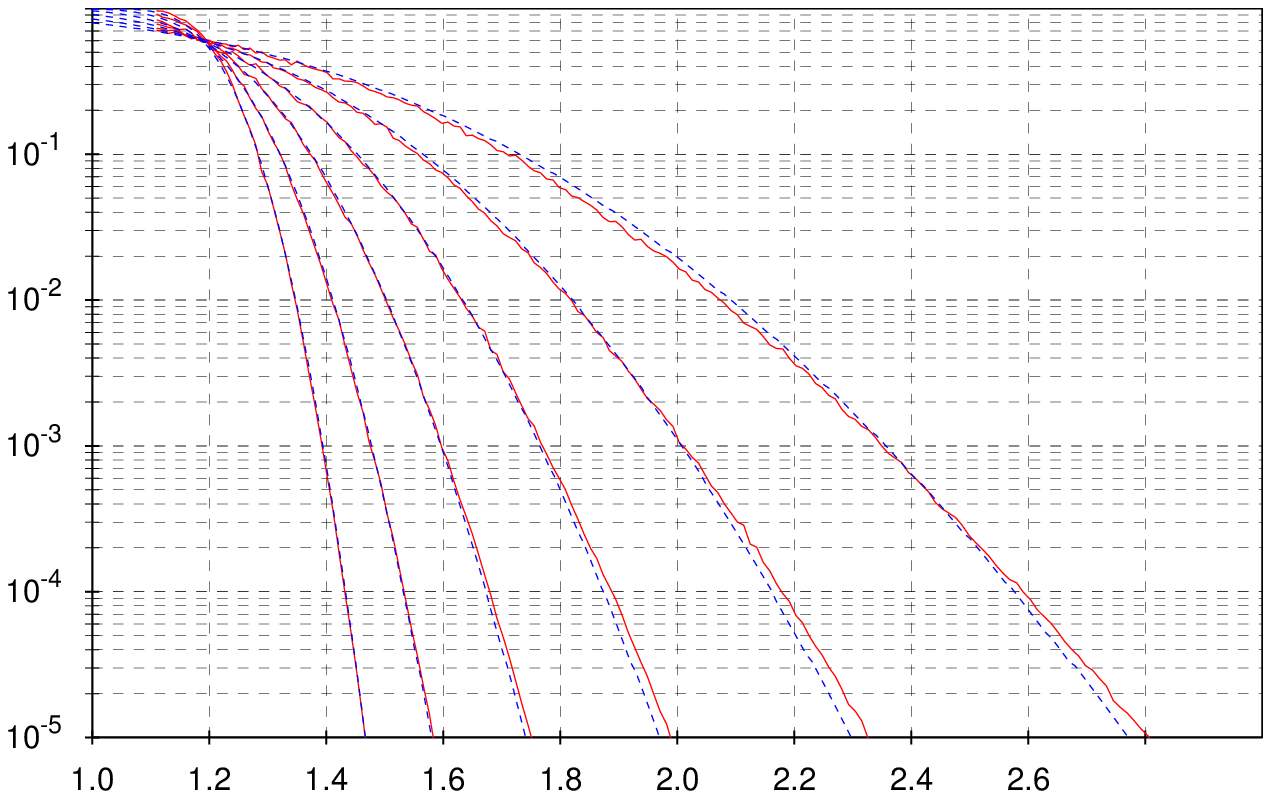}
\end{picture}%
\begin{picture}(367.5,231)
\put(13,223){\makebox(0,0){$\prob_{\text{B}}$}}
\put(380,8){\makebox(0,0)[r]{$\left( E_b/N_0 \right)_{\text{dB}}$}}
\end{picture}
\end{center}
\caption{\label{fig:scalingawgnquant}
Scaling of $\prob_{\rm B}(n, \sigma)$ for transmission over 
BAWGNC$(\sigma)$ and a quantized version of belief propagation decoding 
implemented in hardware.  The threshold for this combination is 
$\left(E_b/N_0 \right)^*_{\text{dB}} \approx 1.19658$.
The blocklengths $n$ are
$n=1000$, $2000$, $4000$, $8000$, $16000$ and $32000$, respectively.
The solid curves represent the simulated ensemble averages.
The dashed curves are computed according to the
refined scaling law (\ref{equ:RefScalingBlock}) 
with scaling parameters $\alpha=0.8694$ and $\beta=5.884$. 
These parameters were fitted to the empirical data.}
\end{figure}
It is worth making a couple of remarks. First of all, the form 
(\ref{equ:generalfss}), wherever it can be argued to hold, allows
a precise definition of what is meant by ``waterfall'' region. 
This is going to be the interval of channel parameters 
$\eit-C_-n^{-\frac{1}{\nu}}\le \epsilon\le \eit+C_+n^{-\frac{1}{\nu}}$
for some positive constants $C_-$ and $C_+$.
Second, even in cases in which the function $f(z)$ cannot be determined
analytically, the statement (\ref{equ:generalfss}) is highly informative,
since it reduces a two-variable function to a single-variable one. Moreover,
in several cases, $f(z)$ can be efficiently given in 
terms of a few parameters. This opens the way to empirical applications of 
Eq.~(\ref{equ:generalfss}). An example is provided in 
Fig.~(\ref{fig:scalingawgnquant}).
\vspace{0.2cm}

Finite-length optimization of code ensembles is an issue of great 
practical relevance. We think that the scaling description 
(\ref{equ:generalfss})-(\ref{equ:generalreffss}) may be an important step 
towards a mathematically well-founded solution of this task.

A first numerical investigation of finite-size scaling for LDPC codes
was presented in Ref.~\cite{Mon01}. Earlier accounts of the present work
appeared in \cite{AMRU03,AMRU04}, and a complete version in \cite{AMRU04b}.
Related ideas were put forward by Lee and Blahut \cite{LeBICC03}
and Zemor and Cohen \cite{ZeC95}.
%
%**********************************************************************
%
\section{Heuristic arguments}

In this paper we consider standard LDPC$(n,\ledge,\redge)$ ensembles.
Here $n$ is the blocklength, and $\ledge$ and $\redge$ denote
the degree distribution of (respectively) variable and check nodes from an 
edge perspective. For the sake of simplicity, we shall often 
refer to {\em regular} ensembles with left degree $l$, and right degree $k$.
The corresponding degree distributions read $\lambda(x) = x^l$ and
$\rho(x) = x^k$.

In order to analyze the iterative decoder, we adopt the point of view
introduced by Luby et al. in \cite{LMSSS97,LMSS01}. 
According to this description, the algorithm proceeds as follows 
(we assume the all-zeros codeword to be transmitted and describe the 
action of the algorithm on the Tanner graph).
Given the received message, the decoder deletes all received variable 
nodes and their incident edges.
In this way one arrives at a {\em residual} graph.  
The decoder proceeds now in an iterative
fashion. If the residual graph contains no degree-one check nodes the 
decoding process stops. Otherwise, the decoder randomly chooses one such 
degree-one check node and deletes it together with the corresponding variable 
node and all its incident edges. In this way a new residual graph results
and a new iteration starts. Decoding is successful if all the graph gets 
deleted by this procedure. In the opposite case, the decoder 
gets stucked in a stopping set.

The state of the algorithm after a fixed number of iterations can be
entirely described by a finite set of integers, providing the number of
variable and check nodes of a given degree. Let us denote this vector 
of integers by $\ux$. For regular ensembles
the situation is even simpler: it is enough to specify the total 
number of variable nodes $v$, the number of  degree-one check nodes
$s$, and the number of check nodes of higher degree $t$.
Therefore, in this case $\ux=(v,s,t)$. Notice that, when decoding starts,
these variables are of order $\Theta(n)$. Each iteration of the decoding 
procedure described above, amounts to a finite increment (or decrement)
in these variables. In particular, for the regular case:
$v$ decreases by one; $s$ decreases by one (the check node chosen at that 
iteration) and increases by the number of degree-two check nodes
which are neighbors of the newly deleted variable; $t$ decreases 
by this last quantity. It is easy to realize that the probability distribution 
of these increments (decrements) depends upon $v,s,t$ only on the scale $n$.
More precisely the probability of a variation $(\Delta v,\Delta s, \Delta t)$
is (up to $1/n$ corrections) a smooth function of $v/n$, $s/n$, 
and $t/n$.

\begin{figure}[htp]
\begin{center}
\epsfig{file=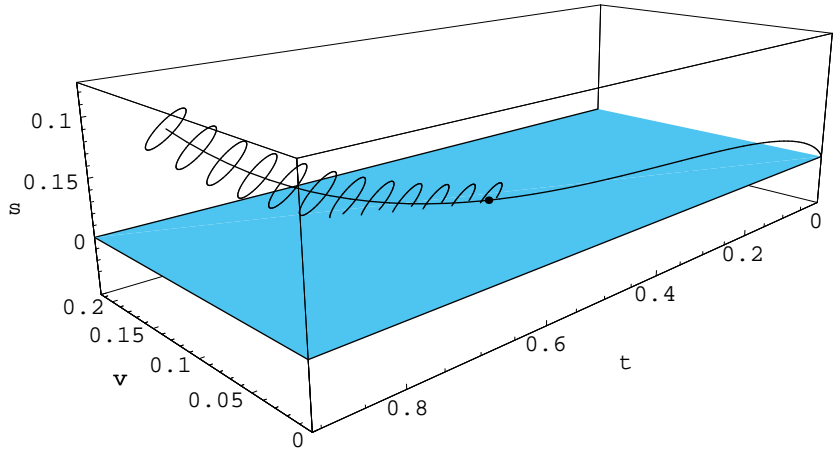, scale=0.9}
\end{center}
\caption{A pictorial representation of density and covariance evolution
for the LDPC($n, x^2, x^5)$ ensemble. Notice that the ellipsoids
corresponding to $(s,t)$ covariances should be regarded as living on a smaller
(by a factor $\sqrt{n}$) scale than the typical trajectory.}
\label{fig:covarianceevolution}
\end{figure}
Call $\ux(\ell)$ the state after $\ell$ iterations, and  
consider the change in state $\delta \ux = \ux(\ell+\delta\ell)-
\ux(\ell)$ in a time $\delta\ell$. If $\delta\ell$ is much smaller
than $\Theta(n)$, then also $|\delta \ux|\ll n$.
Therefore each step in the interval $[\ell,\ell+\delta\ell]$
is independent and 
identically distributed. If $\delta\ell$ is nevertheless much larger than 1, 
we can apply the central limit theorem to deduce that $\delta\ux$
is, with good approximation, a multi-dimensional gaussian variable
with mean of order $\delta\ell$ and standard deviation of order 
$\sqrt{\delta \ell}$. Since, for $\delta\ell\gg 1$,
$\sqrt{\delta \ell}\ll \delta\ell$, one can, to a first approximation
neglect fluctuations. This was the essential step taken in 
\cite{LMSSS97,LMSS01}. These authors showed that $\ux(\ell)$
concentrates around its average value  $\ux_{\rm av}(\ell)\approx 
n\uz(\ell/n)$, with $\uz(\tau)$ solution of the a set of ordinary differential
equations:
\begin{eqnarray}
\frac{d z_i}{d\tau} = f_i(\uz,\tau)\, .
\end{eqnarray}
These are nothing but the density evolution equations. They are integrated with
an initial condition depending upon the erasure probability.

A typical decoding trajectory is reported in 
Fig.~\ref{fig:covarianceevolution}.  For $\epsilon<\eit$, it reaches the 
$(v=0,s=0,t=0)$ point before touching the $s=0$ plane: decoding is successful.
For $\epsilon>\eit$, it touches the $s=0$ plane before the  $(v=0,s=0,t=0)$ 
point: a stopping set has been reached.

Once the typical trajectory is found,
one can compute distribution of $\ux(\ell)-\ux_{\rm av}(\ell)$. The
procedure is conceptually simple. Consider $\ell \approx n\tau $
for some fixed $\tau$ and decompose the interval $[0,\ell]$ into 
sub-intervals of size $\delta\ell$, with $1\ll \delta\ell\ll n$.
Within each sub-interval we can apply the argument outlined above 
to show that $\delta\ux$ is gaussian with standard deviation 
of order $\sqrt{\delta\ell}$. Gluing together $n/\delta\ell$ such intervals
we deduce that  $\ux(\ell)-\ux_{\rm av}(\ell)$ is gaussian with standard 
deviation of order $\sqrt{n}$. This has immediate implications for the
decoding performances. In fact, as soon as the average trajectory 
passes within a distance of order 
$\sqrt{n}$ {\em above} the $s=0$ plane, fluctuations will produce
a finite probability of hitting the plane. Vice-versa, if the average 
trajectory passes $\sqrt{n}$ {\em below} the $s=0$ plane, decoding can 
nevertheless be successful with finite probability.
Recall that the erasure probability sets the initial condition. 
It is easy to realize that a change $\delta\epsilon$ in the channel parameter 
implies a change of order $n\,\delta\epsilon$ in the average trajectory
$\ux_{\rm av}(\ell)$. At threshold ($\epsilon=\eit$),  $\ux(\ell)$ is
just tangent to the $s=0$ plane. This
implies that the failure probability is strictly between $0$ and $1$
as long as $n|\epsilon-\eit|\lesssim \sqrt{n}$. This suggest that
the scaling form (\ref{equ:generalfss})  holds with $\nu=2$.

The above argument can be made both quantitative and rigorous:
the procedure for computing gaussian fluctuations around the typical 
trajectory has been named {\em covariance evolution}~\cite{AMRU04b} 
and it is not much
harder than usual {\em density evolution}. We refer to the next section
for a more precise account of the results. Unhappily, when compared 
with detailed simulations, or with exact computations obtained 
from recursion relations, the results are not very accurate. The reason 
lies in some subtle effect  that produces sizeable corrections of the
form (\ref{equ:generalreffss}). It turns out that $\omega=1/6$: this 
implies large corrections even for quite large blocklengths 
($n$ in the range  $10^3\div 10^5$). Since it turns out that 
$g(z)\propto f'(z)$, the same corrections 
can be attributed to a {\em finite-length shift} of the iterative threshold:
\begin{eqnarray}
\eit(n) = \eit - \eit_1\, n^{-2/3} +\Theta(n^{-1})\, ,
\label{equ:Shift}
\end{eqnarray}
with a positive (ensemble-dependent) constant $\eit_1$.
\begin{figure}[htp]
\begin{center}
\setlength{\unitlength}{1bp}%
\begin{picture}(0,0)
\epsfig{file=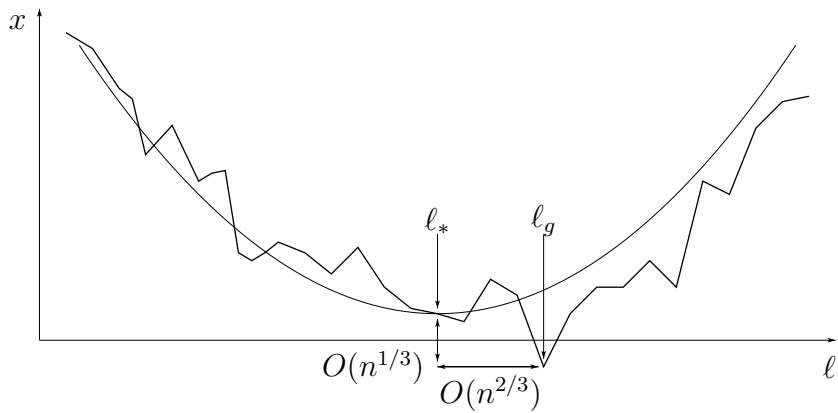,scale=1.0}
\end{picture}%
\begin{picture}(310, 160)
\put(160,75){\makebox(0,0){$\ell_*$}}
\put(200,75){\makebox(0,0){$\ell_g$}}
\put(180,10){\makebox(0,0){$O(n^{2/3})$}}
\put(155,20){\makebox(0,0)[r]{$O(n^{1/3})$}}
\put(310,20){\makebox(0,0)[r]{$\ell$}}
\put(5,150){\makebox(0,0)[r]{$x$}}
\end{picture}
\end{center}
\caption{A pictorial view of decoding trajectories near the critical
point. The type of trajectory depicted here is responsible for 
the finite-length shift of the iterative threshold 
(\ref{equ:Shift}).}
\label{fig:pictorial}
\end{figure}

It is not hard to understand the origin of the shift (\ref{equ:Shift})
heuristically. In a nutshell: when passing $\sqrt{n}$ above the 
$s=0$ plane, the decoding trajectory has many occasions to fail. 
At any time, a fluctuation may imply it touching the plane. On the other  
hand, for decoding to be successful, all fluctuations must be lucky enough
to keep the trajectory away from the plane. This asymmetry leads to
a finite-size lowering of the threshold.

In order to understand the $-2/3$ exponent in Eq.~(\ref{equ:Shift}),
it may be convenient to consider a toy example, cf. Fig.~\ref{fig:pictorial}.
Here the state is describe by a single integer $x$, playing the role of 
$s$ in the decoding problem, evolving over discrete time $\ell$.
Both $x$ and $\ell$ are typically of order $n$, and the increment 
probabilities of $x$ depends smoothly on $x/n$ and $\ell/n$. Finally,
the average trajectory $x_{\rm av}(\ell)$ has a minimum at $\ell_*$,
with $x_{\rm av}(\ell_*) = 0$, and
\begin{eqnarray}
x_{\rm av}(\ell) = \frac{1}{2n}(\ell-\ell_*)^2 +\Theta((\ell-\ell_*)^3/n^2)\, ,
\end{eqnarray}
near the minimum. In other words the minimum is `non-degenerate'.
This is the situation for iterative decoding at $\epsilon = \eit$
under mild conditions on the code ensemble.  
We want to compute the `failure' probability,
$\prob(n)$ i.e. the probability for
the trajectory to touch the $x=0$ plane, which corresponds to the block error
probability in the decoding problem.

Within a first approximation $x(\ell)$ is at any time a gaussian variable
with mean $x_{\rm av}(\ell)$ and standard deviation of order $\sqrt{n}$. 
The failure probability can be estimated
as the probability for $x(\ell= \ell_*)\le 0$. 
Since $x_{\rm av}(\ell= \ell_*) = 0$, we get $\prob(n) =1/2$.
 The crucial
point is now that, even if $x(\ell_*)>0$, the trajectory has some 
probability for touching the $x=0$ plane either for $\ell<\ell_*$
or for $\ell>\ell_*$. What matters is clearly the location of
the minimum $\ell_g$. The trajectory does not touch the $x=0$
plane if and only if $x(\ell_g)>0$.

Let us now try to estimate the position of the minimum $\ell_g$. This is the
outcome of a balance between two competing forces. On the one hand,
the average trajectory is bent upward and forces $\ell_g$ to be close to
$\ell_*$. This yields a contribution to $x(\ell)-x(\ell_*)$ which is of
order $(\ell-\ell_*)^2/2n$. On the other hand, by moving away from
$\ell_*$ one can take advantage of fluctuations,  
$[x(\ell)-x(\ell_*)]-[x_{\rm av}(\ell)-x_{\rm av}(\ell_*)]$
which are  typically of order $\sqrt{\ell-\ell_*}$.
The  location of $\ell_g$ is estimated by balancing these two effects:
$(\ell-\ell_*)^2/2n\sim\sqrt{\ell-\ell_*}$. We get therefore
\begin{eqnarray}
|\ell_g-\ell_*| = O(n^{2/3})\, ,\;\;\;\;\; |x(\ell_g)-x(\ell_*)| = 
O(n^{1/3})\, .
\end{eqnarray}

The above argument implies that the failure probability is slightly larger
than $1/2$. One can in fact fail either if $x(\ell_*)<0$ 
(and this happens with probability $1/2$), or if $x(\ell_*)=O(n^{1/3})$
because, in this case, $x(\ell_g)$ can be $O(n^{1/3})$ below $x(\ell_*)$.
What is the probability of  $x(\ell_*)=O(n^{1/3})$? We know that
 $x(\ell_*)$ is, with good approximation a gaussian variable with mean $0$
and standard deviation $\Theta(\sqrt{n})$. Therefore the probability
is of order $n^{-1/2}\cdot n^{1/3} = n^{-1/6}$. We find therefore the estimate
\begin{eqnarray}
\prob(n) = \frac{1}{2} + \prob_1\, n^{-1/6}+\dots\, .
\end{eqnarray}
Remarkably, the constant $\prob_1$ can be calculated exactly~\cite{AMRU04b}
and depends
uniquely upon the transition probabilities near $x(\ell_*)$. Once
the axes $x$ and $\ell$ have been properly scaled $\prob_1$ is given by an 
integral expression in terms of Airy functions.

When adapted to the coding problem, the above argument yields $\omega=1/6$ 
in Eq.~(\ref{equ:generalreffss}). If we define the finite-size threshold 
$\eit(n)$ by $\block(n,\eit(n)) =1/2$, we get Eq.~(\ref{equ:Shift}).
%
%**********************************************************************
%
\section{Results}

The heuristic arguments presented in the previous Section can be put on 
precise quantitative bases. For some of them we were able to provide
a rigorous foundation, leading to the following:
\begin{lemma}\rm [Scaling of Unconditionally Stable Ensembles]
\label{UnconditionalLemma}
Consider transmission over the BEC$(\epsilon)$ using random elements from
an ensemble $\eldpc n {\ledge} {\redge}$
which has a single critical point and is unconditionally stable. 
Let $\epsilon^*=\epsilon^*(\ledge, \redge)$ denote the threshold
and let $\nu^*$ denote the fractional size of the residual graph at the critical point corresponding to the threshold.
Fix $z$ to be $z:=\sqrt{n}(\epsilon^*-\epsilon)$. 
Let $\prob_{\rm{b}}(n, \ledge, \redge, \epsilon)$
denote the expected bit erasure probability and let 
$\prob_{\rm{B},\gamma}(n, \ledge, \redge, \epsilon)$
denote the expected block erasure probability {\em due to errors of size
at least $\gamma \nu^*$}, where $\gamma \in (0, 1)$.
Then as $n$ tends to infinity,
\begin{eqnarray}
\prob_{\rm{B},\gamma}(n, \ledge, \redge, \epsilon) & = &Q\left(\frac{z}{\alpha}\right) (1+o_n(1)), \label{equ:ScalingBlock}\\
\prob_{\rm{b}}(n, \ledge, \redge, \epsilon) & = &\nu^* Q\left(\frac{z}{\alpha}\right) (1+o_n(1)),\label{equ:ScalingBit}
\end{eqnarray}
where $\alpha=\alpha(\ledge, \redge)$
is a constant which depends on the ensemble. 
\end{lemma}

Unhappily, we were not able to derive powerful enough estimate for
making the `shift' argument rigorous. However, the difficulty is more 
technical than conceptual. We formulate therefore the following:
\begin{conjecture}\rm [Refined Scaling of Unconditionally Stable Ensembles]
\label{RefinedConjecture}
Consider transmission over the BEC$(\epsilon)$ using random elements from
an ensemble $\eldpc n {\ledge} {\redge}$
which has a single critical point and is unconditionally stable.
Let $\epsilon^*=\epsilon^*(\ledge, \redge)$ denote the threshold
and let $\nu^*$ denote the fractional size of the residual graph at the threshold.
Let $\prob_{\rm{b}}(n, \ledge, \redge, \epsilon)$
denote the expected bit erasure probability and let
$\prob_{\rm{B}, \gamma}(n, \ledge, \redge, \epsilon)$
denote the expected block erasure probability {\em due to errors of size
at least $\gamma \nu^*$}, where $\gamma \in (0, 1)$.
Fix $z$ to be $z:=\sqrt{n} (\epsilon^*-\beta n^{-\frac{2}{3}}-\epsilon)$.
Then as $n$ tends to infinity,
\begin{eqnarray}
\prob_{\rm{B}, \gamma}(n, \ledge, \redge, \epsilon)
& = & Q\left(\frac{z}{\alpha} \right)\left(1+O(n^{-1/3}) \right), 
\label{equ:RefScalingBlock}\\
\prob_{\rm{b}}(n, \ledge, \redge, \epsilon)
& = &\nu^* Q\left(\frac{z}{\alpha} \right)\left(1+O(n^{-1/3}) \right),
\label{equ:RefScalingBit}
\end{eqnarray}
where $\alpha=\alpha(\ledge, \redge)$ and $\beta=\beta(\ledge, \redge)$
are constants which depend on the ensemble.
\end{conjecture}

\begin{figure}[htp]
\begin{center}
\setlength{\unitlength}{0.8bp}%
\begin{picture}(0,0)
\includegraphics[scale=0.8]{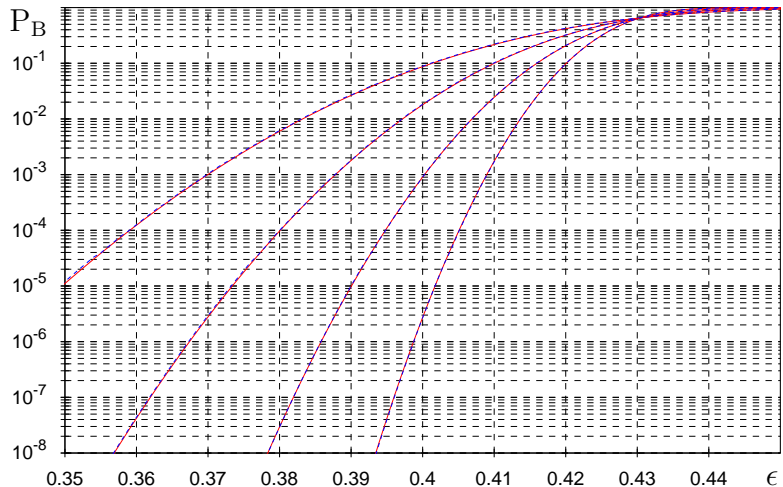}
\end{picture}%
\begin{picture}(367.5,231)
\put(13,223){\makebox(0,0){$\prob_{\text{B}}$}}
\put(366,8){\makebox(0,0)[r]{$\epsilon$}}
\end{picture}
\end{center}
\caption{\label{fig:scalingbec}
Scaling of $\prob_{\rm B}(n, \epsilon)$ for
transmission over BEC$(\epsilon)$ and belief propagation decoding.
The threshold for this combination is $\epsilon^* \approx
 0.42944$, see Table \ref{tab:ParametersRegular}.
The blocklengths/expurgation parameters are 
$n/s=1024/24$, $2048/43$, $4096/82$ and $8192/147$, respectively.
(More precisely, we assume that the ensembles
have been expurgated so that graphs in this ensemble do not contain stopping sets of size $s$ or smaller.)
The solid curves represent the exact ensemble averages.
The dashed curves are computed according to the
refined scaling law stated in Conjecture \ref{RefinedConjecture} with scaling parameters 
$\alpha=\sqrt{0.249869^2+\epsilon^*(1-\epsilon^*)}$ 
and $\beta=0.616045$, see Table \ref{tab:ParametersRegular}.}
\end{figure}

In Table \ref{tab:ParametersRegular} we report the values of the scaling 
parameters for a few regular ensembles. These parameters  are obtained 
by integrating a  set of ordinary differential (covariance evolution)
equations and are easily pushed to great precision.
\begin{table}
\begin{center}
\begin{tabular}{|c|c||c|c|c|} \hline
$l$ & $k$ & $\epsilon^*$ & $\alpha$ & $\beta/\Omega$  \\ \hline \hline
3 & 4 & $0.6473$ & $0.260115$ & $0.593632$ \\ \hline
3 & 5 & $0.5176$ & $0.263814$ & $0.616196$ \\ \hline
3 & 6 & $0.4294$ & $0.249869$ & $0.616949$ \\ \hline
4 & 5 & $0.6001$ & $0.241125$ & $0.571617$ \\ \hline
4 & 6 & $0.5061$ & $0.246776$ & $0.574356$ \\ \hline
5 & 6 & $0.5510$ & $0.228362$ & $0.559688$ \\ \hline
6 & 7 & $0.5079$ & $0.280781$ & $0.547797$ \\ \hline
6 & 12 & $0.3075$ & $0.170218$ & $0.506326$ \\ \hline
\end{tabular}
\label{tab:ParametersRegular}
\end{center}
\caption{Thresholds and scaling parameters for some regular standard ensembles
$\lambda(x) = x^l$, $\rho(x) = x^k$.
The shift parameter is given
as $\beta/\Omega$ where $\Omega$ is the universal constant defined in 
Ref.~\cite{AMRU04b} in terms of Airy functions, and 
whose numerical value is very close to 1.}
\end{table}
Finally, in Fig. \ref{fig:scalingbec} we compare the refined scaling 
form provided by Conjecture 1, with the exact error probability computed by 
recursion. The agreement is excellent.

\section*{Acknowledgments}

The work of AM was partially supported by European Community 
under the project EVERGROW.
%
%**********************************************************************
%

\end{document}